\documentclass[12pt,preprint]{aastex}
\usepackage{graphicx, amsmath, amssymb}
\usepackage[dvips]{color}

\newcommand{\pow}[2]{\ensuremath{\mbox{#1}^{\rm #2}}}
\newcommand{\subs}[2]{\ensuremath{\mbox{#1}_{\rm #2}}}

\newcommand{\Blos}{\ensuremath{B_{\rm los}}}

\newcommand{\kmS}{km \pow{s}{-1}}
\newcommand{\cm}[1]{\pow{cm}{#1}}    

\newcommand{\vLSR}{\ensuremath{v_\text{LSR}}}
\newcommand{\beam}{\pow{beam}{-1}}
\newcommand{\Texct}{\ensuremath{T_{\rm ex}}}
\newcommand{\Av}{\subs{A}{v}}

\shortauthors{Sarma, et al.}
\shorttitle{VLA OH Zeeman observations of S88B}

\begin{document}

\title{VLA OH Zeeman Observations of the star forming region S88B}

\author{A.~P.~Sarma\altaffilmark{1},
C.~L.~Brogan\altaffilmark{2},
T.~L.~Bourke\altaffilmark{3},
M.~Eftimova\altaffilmark{1,4},
T.~H.~Troland\altaffilmark{5}}

\altaffiltext{1}{Physics Department, DePaul University, 
2219 N. Kenmore Ave., Byrne Hall 211, 
Chicago IL 60614; asarma@depaul.edu}

\altaffiltext{2}{National Radio Astronomy Observatory, Charlottesville, VA 22903}

\altaffiltext{3}{Harvard-Smithsonian Center for Astrophysics, Cambridge, MA 02138}

\altaffiltext{4}{Current Address: Dept.\ of Physics \& Astronomy, 
University of North Carolina, Chapel Hill, NC 27599}

\altaffiltext{5}{Dept.\ of Physics \& Astronomy, University of Kentucky, Lexington KY 40506}

\begin{abstract}

We present observations of the Zeeman effect in OH thermal
absorption main lines at 1665 and 1667 MHz taken with the 
Karl G. Jansky Very Large Array (VLA) toward the star forming region S88B. 
The OH absorption profiles toward this source are complicated, and
contain several blended components
toward a number of positions. Almost all of the OH absorbing
gas is located in the eastern parts of S88B, toward the compact
continuum source S88B-2 and the eastern parts of the extended
continuum source S88B-1. The ratio of 1665/1667 MHz OH line intensities 
indicates the gas is likely highly clumped, in agreement with
other molecular emission line observations in the literature.
S88-B appears to present a similar geometry to the well-known
star forming region M17, in that there is an edge-on eastward progression
from ionized to molecular gas. The detected magnetic fields appear
to mirror this eastward transition; we detected line-of-sight magnetic
fields ranging from 90-400 $\mu$G, with the lowest values of
the field to the southwest of the S88B-1 continuum peak,
and the highest values to its northeast. We used the detected fields
to assess the importance of the magnetic field in S88B by a number
of methods; we calculated the ratio of thermal to magnetic pressures,
we calculated the critical field necessary to completely support the cloud 
against self-gravity and compared it to the observed field, and we
calculated the ratio of mass to magnetic flux in terms of the critical value
of this parameter. All these methods indicated that the magnetic field in
S88B is dynamically significant, and should provide an important source
of support against gravity. Moreover, the magnetic energy density is in approximate
equipartition with the turbulent energy density, again pointing to the 
importance of the magnetic field in this region.
\end{abstract}


\section{INTRODUCTION}
\label{sINTRO}

Massive stars affect the structure and dynamics of galaxies, the planet
formation process, and even the formation of low mass stars via 
outflows, winds, and supernova explosions that drive mixing and
turbulence in the interstellar medium (e.g., \citealt{zy07}, 
and references therein). Yet, our understanding of their formation
and evolution remains unclear. Moreover, the importance of 
magnetic fields in the formation of stars, both low and high mass,
has long been acknowledged (e.g., \citealt{mou87}; \citealt{hgmz93};  
\citealt{rmc99}; \citealt{mck99}). Indeed, ongoing theoretical
work continues to probe the effects of magnetic fields on the
star formation process (e.g., \citealt{zhao2011}; \citealt{meli2009}; \citealt{boss07}; \citealt{shu06}; 
\citealt{mtk06}; \citealt{naka05}). However, observational data
on magnetic fields is still scarce and theoretical work is
hampered by the lack of information on magnetic field 
strengths in different density regimes.

With these considerations in view, we have observed the high mass
star forming region S88B with the Karl G. Jansky Very Large Array (VLA) for the Zeeman
effect in thermal absorption lines of OH at 1665 and 1667 MHz. 
Observations of the Zeeman effect in absorption lines with interferometers
like the VLA provide an excellent method of $mapping$ the magnetic 
field in regions that are along the line of sight toward strong background 
continuum sources (e.g., \citealt{bt2001}; \citealt{stc2000}). 
In the next subsection, we will present some details of the source
S88B. In \S\ \ref{sODR}, we present details of the observations 
and reduction of the data. The results are presented in \S\ \ref{sR}, and 
discussed in \S\ \ref{sDISC}. In \S\ \ref{sCONC}, we state our conclusions.

\subsection{S88B}
\label{ssISRC}
S88B is a high mass star forming region (near $l = 61\fdg48, b = 0\fdg09$)
with a bolometric luminosity of $9.0 \times 10^4\ L_\Sun$ (\citealt{mse02}). 
It has been observed at a wide variety of wavelengths. There is some confusing
nomenclature in the literature, and so we begin by discussing the 
morphology of its surroundings. S88 is itself a diffuse extended nebula,
and is likely a very evolved \ion{H}{2}\ region (e.g., \citealt{fh81}).
About 14\arcmin\ to its southeast lie two bright knots of optical
nebulosity: S88A and S88B (e.g., see the red Palomar Sky Survey print
in Fig.\ 1 of \citealt{eh81}). No radio continuum emission is observed from the western
source, S88A. Strong radio emission is observed from
an \ion{H}{2}\ region centered slightly to the east of the optical
nebulosity S88B, and this is the target source for our Zeeman observations. 
\citet{fh81}\ first proposed that the radio source in S88B was composed 
of two components, based on their 5 GHz continuum observations. The
more extended component to the west (e.g., see \S\ \ref{ssC}, and Fig.\ 1 
in this paper) that intersects the optical nebulosity
is S88B-1, whereas the compact component to the east is S88B-2. In their
high resolution 6 cm VLA image (HPBW 0.4\arcsec), \cite{wc89}\ resolved
out many details of these two sources --- what they call 
G61.48+0.09A is referenced as S88B-2 in this paper (and most of
the literature), whereas their source G61.48+0.09B is a group of low 
intensity sources to the southwest of S88B-1.
Following \citet{eh81}, we adopt the distance to S88B of 2.0 kpc. 
More recently, \citet{swa04}\ have claimed a distance of 4.1 $\pm$ 2.4
kpc to S88B on the basis of their H110$\alpha$ recombination line and 
H$_2$CO absorption observations; we note that our adopted value is 
within their error limits. Moreover, \citet{ggp03}\ have estimated on 
the basis of foreground star counts that the distance to S88B cannot 
be substantially greater than 2.0 kpc.

\section{OBSERVATIONS \& DATA REDUCTION}
\label{sODR}
The observations were carried out with the Very Large Array (VLA) of the 
NRAO\footnote{The National Radio Astronomy Observatory (NRAO) is a 
facility of the National Science Foundation operated under cooperative 
agreement by Associated Universities, Inc.} in 2003 in the B-configuration,
and combined with C-configuration data observed in 1997. Important 
parameters of the observations are listed in Table\ \ref{tOP}. Both 
circular polarizations and both main lines (1665 and 1667 MHz) of OH
were observed simultaneously in absorption. In order to mitigate instrumental 
effects, a front-end transfer switch was used to alternate the sense of circular 
polarization passing through each telescope's IF system every $\sim$10 
minutes.  

The editing, calibration, Fourier transformation, deconvolution, and 
calculation of optical depths for the OH data were carried out using the 
Astronomical Image Processing System (AIPS) of the NRAO. Following
standard procedure for Zeeman effect observations, bandpass correction
was applied only to the Stokes $I$ data set, since bandpass effects 
subtract out to first order in the Stokes $V$ data, and the bandpass correction process adds noise to the $V$ profiles. The B and C 
configuration data were combined in the uv-plane using the
AIPS task $DBCON$. The rms in the continuum image of the combined
B and C configuration data is 0.8 mJy \beam, whereas the rms in
a channel of the line data is 1.7 mJy \beam. Magnetic fields
were determined as described in \S\ \ref{ssB}\ by using the 
Multichannel Image Reconstruction, Image 
Analysis and Display (MIRIAD) software processing package, formerly of the 
Berkeley-Illinois-Maryland Array (BIMA) and now incorporated into
the Combined Array for Research in Millimeter-Wave Astronomy (CARMA)
telescope.

\section{RESULTS}
\label{sR}

\subsection{The 18 cm Continuum}
\label{ssC}
\begin{figure}	
\centering
\includegraphics[width=0.9\textwidth]{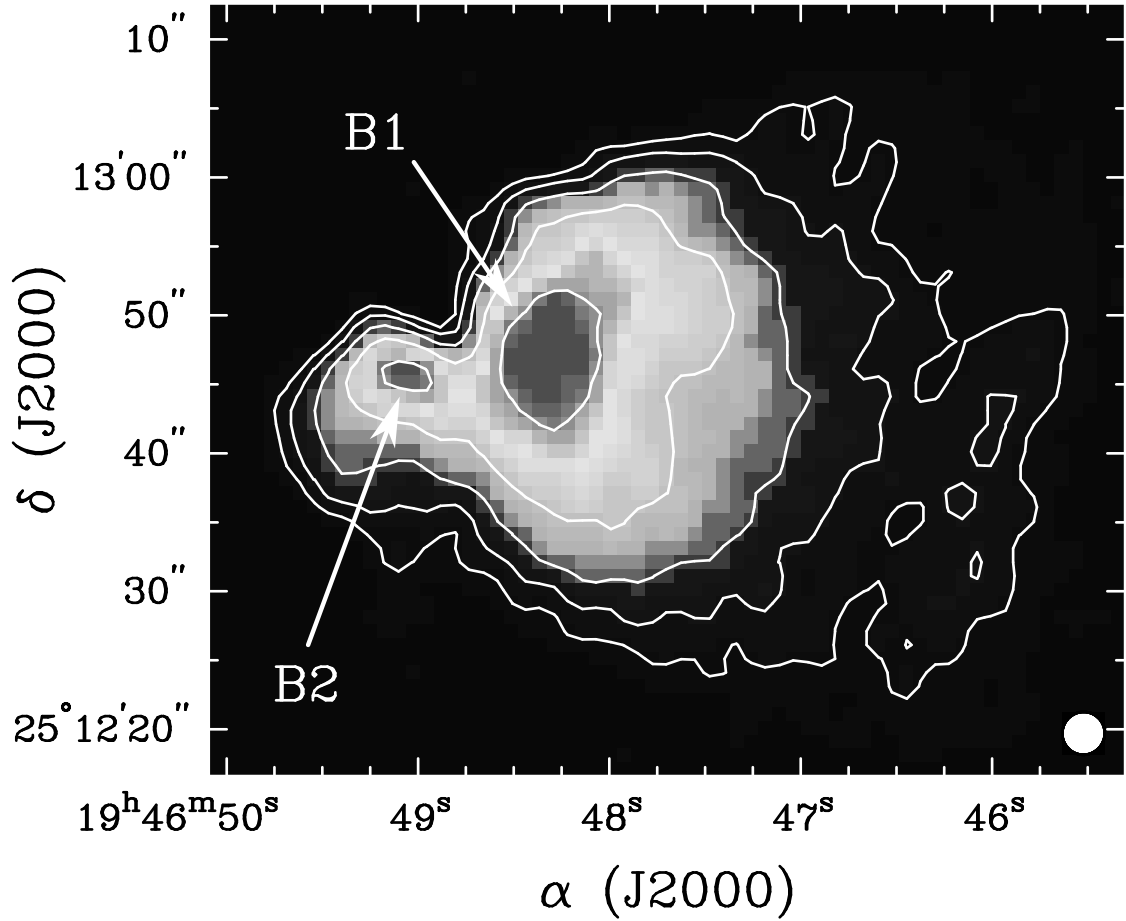}
\caption{Contour image of the 18 cm continuum toward S88B. 
The beam is 3$\farcs$1 $\times$ 3$\farcs$0 (P.A.=$-6\fdg$1), and the 
rms in the image is 0.8 mJy \beam. The contours are 
at 6, 12, 24, 48, 96 mJy \beam.  The extended western component
S88B-1, and the compact component to the east S88B-2, are marked
in the figure as B1 and B2 respectively. \label{fC}}
\end{figure}

Figure\ \ref{fC}\ shows our highest resolution 18 cm continuum image of
S88B made with uniform weighting (beam HPBW $3\farcs1 \times 3\farcs0$). 
The extended western component S88B-1 and the compact component 
on the eastern side S88B-2, first identified by \citet{fh81},
are clearly visible in this figure. We determined the integrated 
flux for S88B-1 to be $4.27 \pm 0.23$ Jy, and the integrated flux for 
S88B-2 to be $0.79 \pm 0.32$ Jy. In a 5 GHz continuum survey with the NRAO 140-ft 
telescope (HPBW 6$\farcm$5), \citet{rw70}\ obtained 6.1 Jy for the
source 61.5+0.1 which is a combination of S88B-1 and S88B-2.
In a 12-hr synthesis map with Westerbork (WSRT) at 5 GHz (HPBW 6$\farcs$6), 
\citet{fh81}\ obtained a total flux density of 6.1 $\pm$ 0.2 Jy for S88B. 
At 4.9 GHz, \citet{grm93}\ measured 4.4 Jy toward S88B-1 and 0.8 Jy 
toward S88B-2 in a combined C and D configuration observation with the 
VLA (HPBW $\sim 4\arcsec$). The integrated flux densities measured with
these shorter wavelength data are in good agreement with the 18 cm
VLA results, suggesting that the emission is predominantly either near 
the turnover point or optically thin ($\alpha=-0.1$, $S_\nu \propto
\nu^\alpha$). Figure\ \ref{fCC}\ shows the 350 $\mu$m dust 
continuum image of \citet{mse02}\ superposed on our 18 cm continuum image, 
along with other pertinent information --- these are discussed 
in \S\ \ref{ssDC}.

\subsection{OH Absorption}
\label{ssH1A}
The OH absorption profiles toward S88B are complicated, with several blended
components toward a number of positions. In order to characterize the velocity 
structure, optical depth profiles for the OH lines were calculated with the AIPS 
task $COMB$ using standard procedure (e.g., see \citealt*{rct95}). The OH 
column density was then determined using the relation
\begin{equation}
N_{\rm OH} / T_{\rm ex} = C \int {{\tau}_v\, \mbox{d}v} 
\phm{abc}\mbox{cm}^{\rm -2}\phm{b}\mbox{K}^{\rm -1} 
\end{equation}
where $T_{\rm ex}$ is the excitation temperature of OH and the 
constant $C = 4.1063 \times 10^{\rm14}$ and $2.2785 \times 10^{\rm14}$ 
$\mbox{cm}^{-2}$ (km $\mbox{s}^{-1}$)$^{-1}$ for the OH 1665 and 
1667 MHz lines respectively (\citealt{c77}). 

The optical depth profiles reveal complex kinematics with a number 
of heavily blended components present toward much of the region. In order 
to highlight the prominent absorption components, contour plots of the OH 
optical depth in several velocity channels are shown in Figure \ref{fTAU}, 
superposed on a grayscale image of the 18 cm continuum. Every third
velocity channel is displayed, and the plots are shown over the velocity
range 17.4 \kmS\ to 24.0 \kmS. Figure\ \ref{fNOH}\ shows plots of 
$N$(OH)/$T_{\rm ex}$ toward S88B at 1667 MHz. In principle, 
variations in $N$(OH)/$T_{\rm ex}$ may arise
due to variations in the OH column density, or due to variations in 
$T_{\rm ex}$, or both. However, in the results and discussion, we
quote only variations in $N$(OH)/$T_{\rm ex}$, since the excitation
temperature cannot be measured based on absorption studies alone. Further
discussion about possible variations in $T_{\rm ex}$ appears in 
\S\ \ref{ssDC}. 

From Figures \ref{fTAU}\ and \ref{fNOH}, we identified four prominent absorption 
enhancements, and these are marked in Figure~\ref{fNOH}\ as OH-a, OH-b, OH-c, 
and OH-d respectively. The absorption enhancement OH-a lies to the southwest of the
continuum peak of S88B-1; it is seen in the 19.1 \kmS\ panel in Figure~\ref{fTAU},
with a peak $\tau \sim 0.5$. OH-a is also seen in the 19.9 \kmS\ panel of 
Figures \ref{fTAU}\ with a higher peak $\tau \sim 1$. In this 19.9 \kmS\ panel,
we also see the enhancements OH-b and OH-c, the former to the northwest of
the continuum peak of S88B-1, and the latter about midway between S88B-1 
and S88B-2. Finally, OH-d is seen in the 20.7, 21.5, and 22.4 \kmS\ panels of
Figure~\ref{fTAU}, with high peak $\tau \sim 2$ in two of these panels; it lies 
northward of the S88B-2 continuum peak; OH-d is also visible at the highest 
velocities at which OH absorption is observed ($\sim 23$ \kmS). The high
$\tau$'s in OH-d account for the peak in $N$(OH)/$T_{\rm ex}$ to the north 
of the continuum peak of S88B-2.

\subsection{Magnetic Field Strengths}
\label{ssB}
The magnetic field strengths in S88B were determined using the Zeeman effect, by fitting a 
numerical frequency derivative of the Stokes $I$ spectrum to the Stokes $V$ 
spectrum for each pixel in the absorption line cube. The Stokes parameters have
the standard definition of $I$ = (RCP + LCP), and $V$ = (RCP $-$ LCP); RCP is right- and 
LCP is left-circular polarization; RCP has the standard radio definition of clockwise rotation of the 
electric vector when viewed along the direction of wave propagation. Since the observed $V$ spectrum 
may also contain a scaled replica of the $I$ spectrum itself, the Stokes $V$ spectral profile 
can be described by the equation (\citealt{th82}; \citealt{skzl90}):
\begin{equation}
V\ = aI\ + \frac{b}{2}\ 
\frac{\mbox{d}I}{\mbox{d}\nu}
\label{eVLCO}
\end{equation}
In this equation, the fit parameter $a$ represents
the scaled-down replica of the $I$ profile in the $V$ profile; $a$
is usually the result of small calibration errors in $RCP$ vs.\ $LCP$;
for all results reported here, $a \le 10^{-3}$. 
For thermally excited lines such as the OH 1665 and 1667 MHz lines
reported in this paper, $b = zB\, \text{cos}\, \theta$, where $B$ is the 
magnetic field, $\theta$ is the angle of the magnetic field to the line of sight, 
and $z$ is the Zeeman splitting factor which is equal to $z$ = 3.27 Hz $\mu$G$^{-1}$ for 
the 1665 MHz line of OH, and 1.96 Hz $\mu$G$^{-1}$ for the 1667 MHz 
line of OH (\citealt{kc86}).  
We determined magnetic field strengths in S88B by using a least- squares method to fit 
equation (\ref{eVLCO}) to the Stokes $V$ spectra to solve for $a$ and $b$.
The results of the fits give the line-of-sight component 
of the magnetic  field, \Blos\ = $B\, \text{cos}\, \theta$.  
It is usual practice to consider the results to be 
significant only if the derived value of \Blos\ is greater 
than  the 3$\sigma$ level. For Zeeman effect observations in OH absorption
lines, we usually impose a stronger condition --- we consider the detections
to be significant only if both 1665 and 1667 MHz \Blos\ values are
greater than the 3$\sigma$ level. In the current observations, however,
this stronger criterion for significance would leave out
some important information on the field. Toward S88B-2, the 1665 MHz 
profiles show a $\sim$200 $\mu$G field with about 5$\sigma$ level of
significance, whereas the 1667 MHz profiles show a field
below the 3$\sigma$ level. Therefore, for S88B-2 only, we have relaxed 
the criterion above to consider the measurement to be significant
if it displays a 3$\sigma$ detection only in the 1665 MHz line. 
We believe this is reasonable, since the measured value of \Blos\ 
toward this source does not appear to be wildly discrepant from the measured 
value toward S88B-1. The non-detection at 1667 MHz toward S88B-2 is not 
surprising. The 1667 MHz Stokes V signal is weaker than the 1665 MHz Stokes V 
signal because the Zeeman splitting factor for OH 1667 MHz
lines ($z$ = 1.96 Hz $\mu$G$^{-1}$) is less than that for OH 
1665 MHz lines ($z$ = 3.27 Hz $\mu$G$^{-1}$). This inequality in 
the Zeeman factors would not impact the result as much if the OH gas was
thermalized and homogenous, since the intensity ratios would then 
be $\tau_{67}/\tau_{65} = 9/5$. However, as described below (\S\ \ref{ssDC}),
$\tau_{67}/\tau_{65} = 1.0-1.3$ toward S88B, which makes the
sensitivity of the 1667 MHz line to the Zeeman effect less than that
of the 1665 MHz line.

The resulting \Blos\ map (taken from the 1665 MHz data) is shown
in Figure\ \ref{fBLOS}. In the area enclosed by the thick-lined ellipse in this
figure, the detected field is above the 3$\sigma$ level only in the
1665 MHz line, whereas in the rest of the displayed \Blos\ map, the 
field is above the 3$\sigma$ level in both 1665 and 1667 MHz lines,
with peak signal-to-noise ratios of $\sim$10.
Examples of the Stokes $I$ and $V$ profiles
together with the derivative of $I$ scaled by the fitted value of \Blos\
are shown in Figure\ \ref{fB1IVD} and Figure\ \ref{fB2IVD}; the locations
toward which these profiles have been plotted are marked in Figure \ref{fBLOS}
by a ``+'' and a ``$\times$'' respectively. From Figure\ \ref{fB1IVD} and Figure\ \ref{fB2IVD}, 
it is evident that the magnetic field is detected in the component with $\vLSR \sim 21~$\kmS; 
no fields were detected in any of the other velocity components.
Toward S88B-1, we see from Figure\ \ref{fBLOS}\ that \Blos\ increases from
about 90 $\mu$G in the southwest to about 400 $\mu$G
in the northeast; this is the case in both 1665 and 1667 MHz lines. Meanwhile,
toward S88B-2, the 1665 MHz line reveals values of \Blos\ as high 
as $\sim$270 $\mu$G.

\section{DISCUSSION}
\label{sDISC}

\subsection{Morphology and Kinematics}
\label{ssDC}
Early optical, radio, and infrared observations of S88B by \citet{psv77}
revealed that the peak of the near infrared (NIR) position is eastward of the 
optical nebula, whereas the radio continuum peak is even farther to the east than the 
NIR peak. This eastward progression is consistent with our observation
of the location of the absorbing OH gas. It is clear from Figure~\ref{fTAU}\ that almost 
all of the OH absorbing gas is located in the eastern half of S88B (i.e., all of
S88B-2, and the eastern part of the extended source S88B-1). We 
did not detect any OH absorption in the western parts of S88B-1, where it is 
coincident with the H$\alpha$ nebulosity. The molecular cloud associated with 
S88B was mapped by \citet{eh81}\ in several CO transitions at low resolution 
(HPBW 1$\arcmin$-2$\arcmin$), and found to extend over 10$\arcmin$ (6.5 pc)
with a total mass of $5 \times 10^3$ M$_\Sun$. \citet{eh81}\ speculated
that the \ion{H}{2}\ regions S88B-1 and S88B-2 are a blister formation 
in the molecular cloud which we are observing from the side. The velocities
of the absorbing OH gas match up well with molecular line velocities;
e.g., the CO lines are singly peaked at 21 \kmS\ (\citealt{eh81}). Moreover,
recombination line observations by \citet{gg94}\ with the VLA 
(HPBW $\sim$ 2$\arcsec$) revealed that the center velocity of the 
single-peaked H92$\alpha$ line emission toward S88B-1 increases from 
22 \kmS\ at the eastern edge of S88B-1 to 32 \kmS\ at the western edge. 
The 22 \kmS\ velocity in the H92$\alpha$ recombination line at the eastern 
edge of S88B-1 matches well with the OH absorption data --- e.g., see the panels
between 21.5 \kmS\ and 22.4 \kmS\ in Figure~\ref{fTAU}. On the other hand, 
the recombination line velocity at the western edge matches well with the 
optical H$\alpha$ velocities which exceed 28 \kmS\ (\citealt{dm78}).
In other words, we are observing an edge-on transition from the ionized
gas to the molecular gas in S88B, similar to M17 (\citealt{bt2001}). 

\begin{figure}	
\centering
\includegraphics[width=0.9\textwidth]{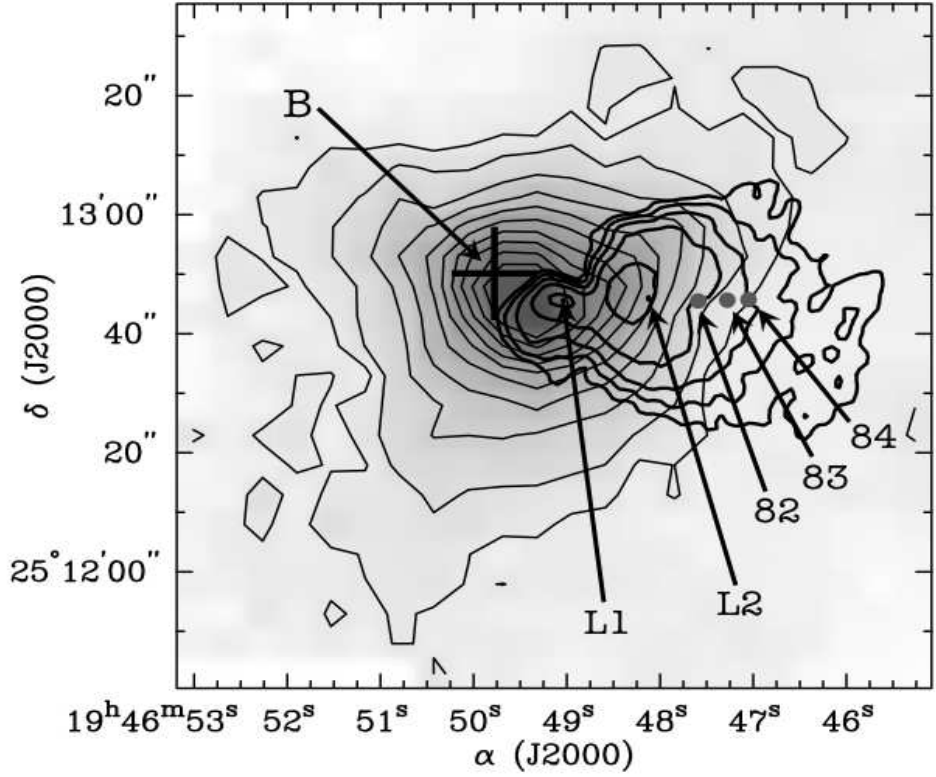}
\caption{Contours and grayscale of the 350 $\mu$m dust emission toward S88B from \citet{mse02}\ 
superposed on the 18 cm continuum image. The HPBW of the
350 $\mu$m data is 14$\arcsec$, and contours are at 1.4, 2 Jy beam$^{-1}$, and then in increments of 1 Jy beam$^{-1}$
up to 10 Jy beam$^{-1}$. Our 18 cm continuum
image has the same resolution and contours as in Figure \ref{fC}.
The positions of the NIR sources 82, 83, and 84 from \citet{dnz2000}, the 3.5 $\mu$m sources L1 and L2 from 
\citet{puga04}, and clump B from the C$^{18}$O observations of \citet{ss07} are also marked in the figure. \label{fCC}}
\end{figure}

\begin{figure}	
\centering
\includegraphics[width=0.9\textwidth,angle=-90]{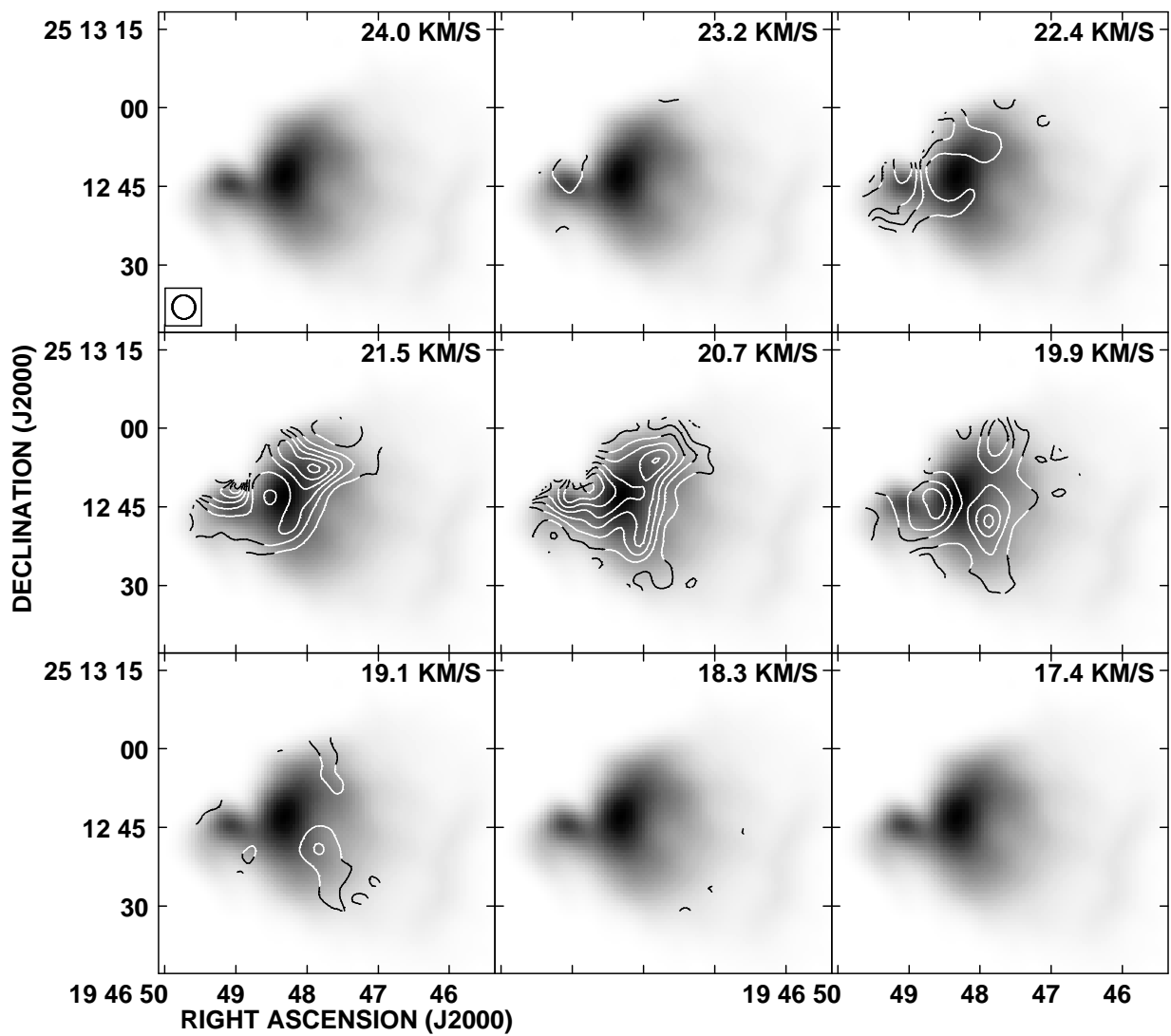}
\caption{OH 1667 MHz optical depth channel images from 17.4 \kmS\ to
24.0 \kmS\ in \vLSR. The contours
are from $\tau$ = 0.25 to $\tau$ = 2.0 in steps of 0.25. Every third
velocity channel is displayed. The grayscale shows the 18 cm continuum
at the same resolution as the OH optical depth image, which is
$4\farcs5 \times 4\farcs3$ (P.A.=$14\fdg1$), as shown in the boxed
inset in the top left panel.\label{fTAU}} 
\end{figure}

\begin{figure}	
\centering
\includegraphics[width=0.9\textwidth]{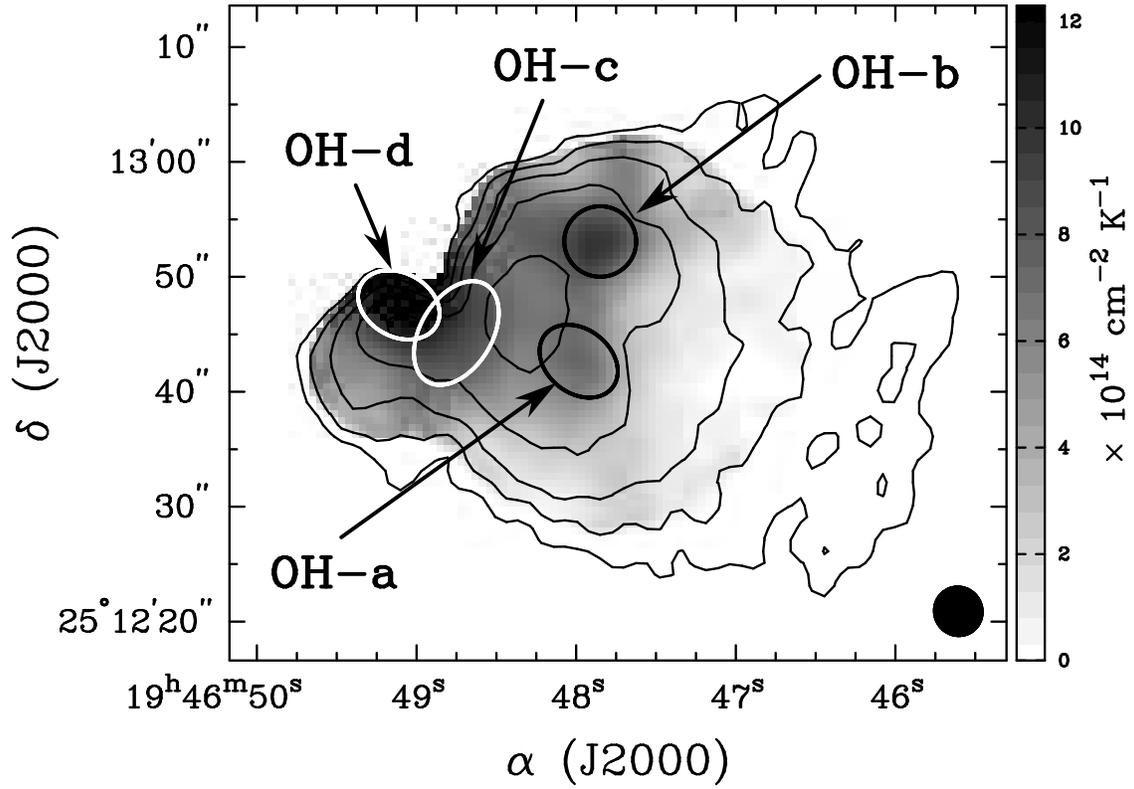}
\caption{Gray-scale image of 
N$_{\rm OH67}$/T$_{\rm ex}$ toward S88B overlaid on the 18 cm 
continuum. The continuum image has the same resolution 
and contours as in Fig.\ \ref{fC}. The 1667 OH beam is shown
in the lower right; its parameters are given in Fig.~\ref{fTAU}.  \label{fNOH}}
\end{figure}

\begin{figure}	
\centering
\includegraphics[width=0.9\textwidth]{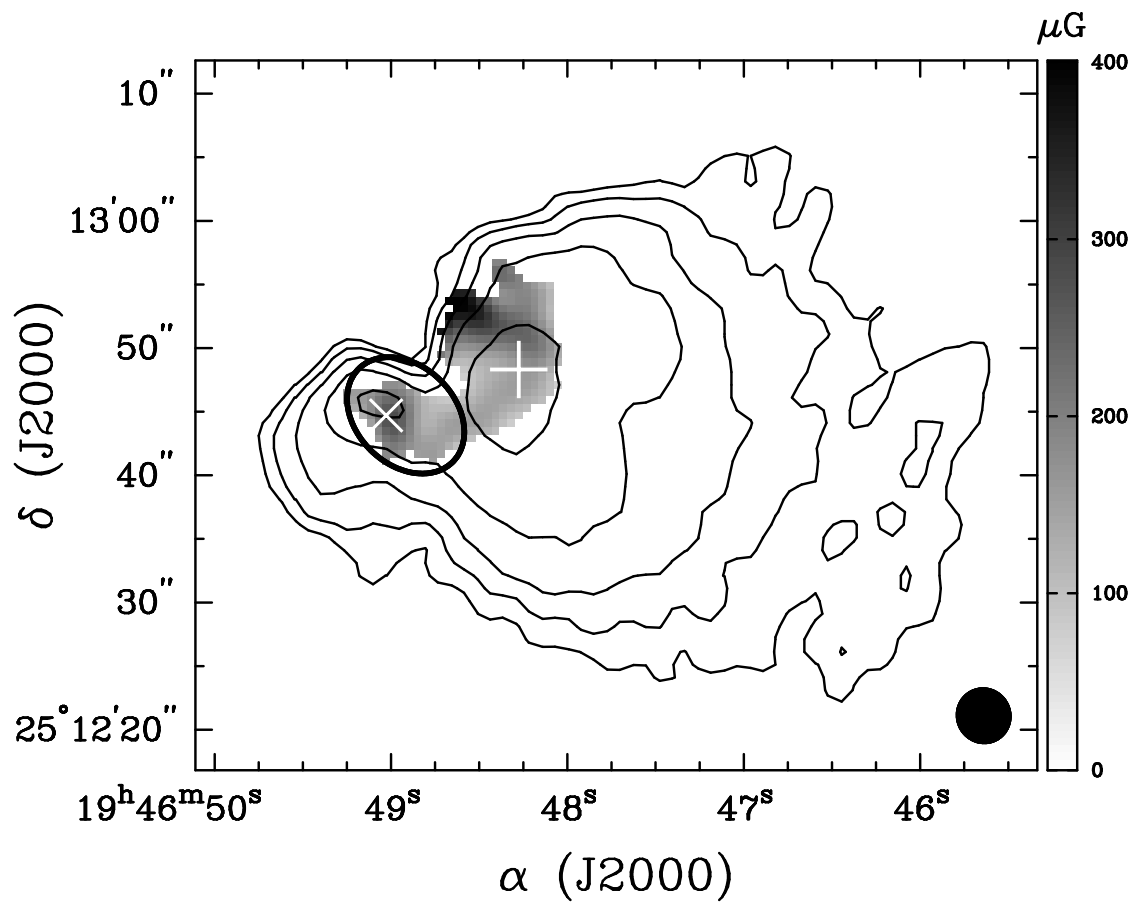}
\caption{Image of the detected \Blos\ toward S88B, taken from the
1665 MHz data. The contours depict the 18 cm continuum, and are 
the same as in Fig\ \ref{fC}. The thick-lined ellipse shows the region in 
which the detected \Blos\ is at the 3-$\sigma$ level in 1665 MHz 
only. The ``+'' sign indicates the position toward which
the profiles are shown in Fig.~\ref{fB1IVD}\ below, while the
``$\times$'' indicates the position toward which
the profiles are shown in Fig.~\ref{fB2IVD}. \label{fBLOS}}
\end{figure}

\begin{figure} 
\centering
\begin{minipage}[t]{0.45\textwidth}
\includegraphics[width=0.9\textwidth]{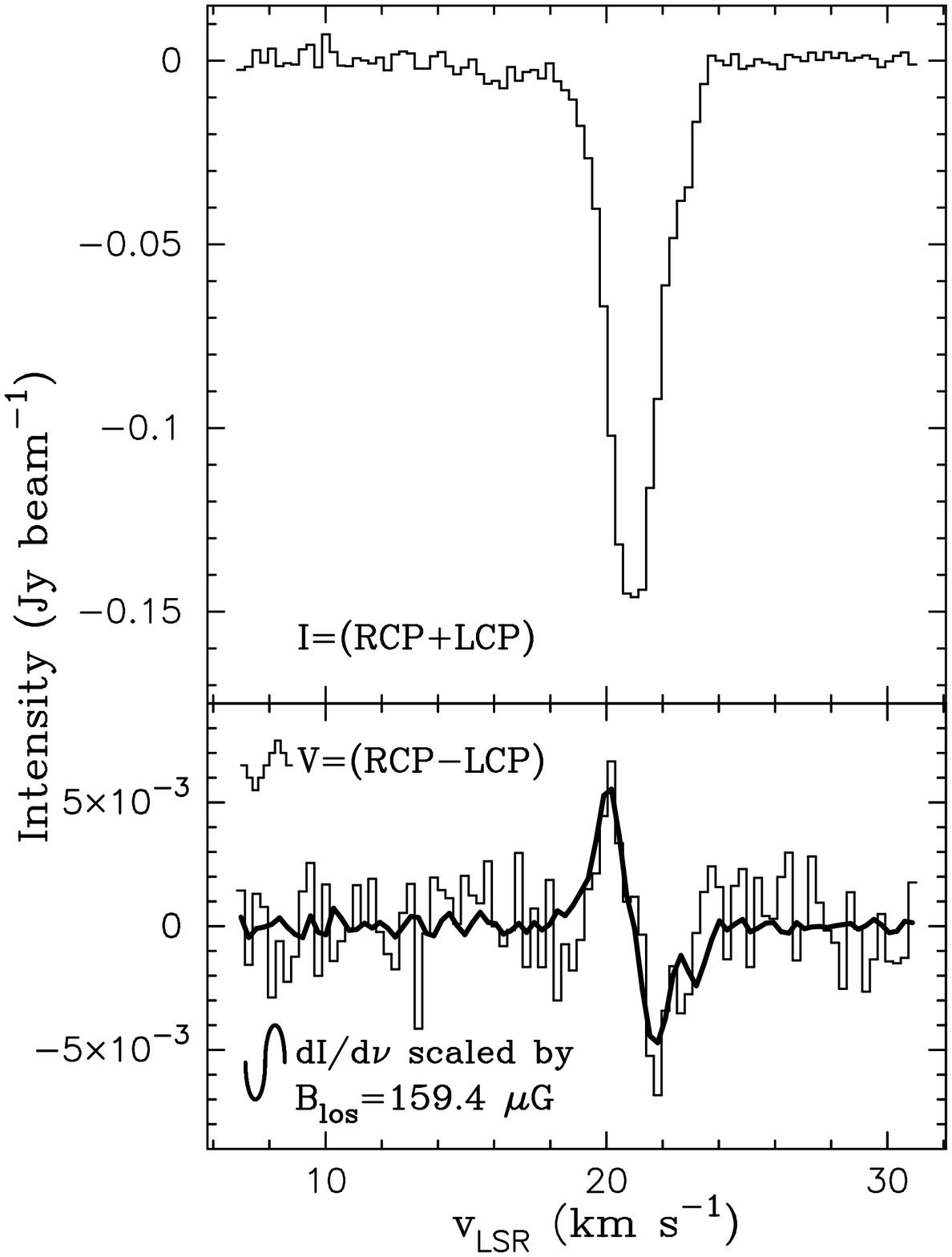}
\end{minipage}
\begin{minipage}[t]{0.45\textwidth}
\includegraphics[width=0.9\textwidth]{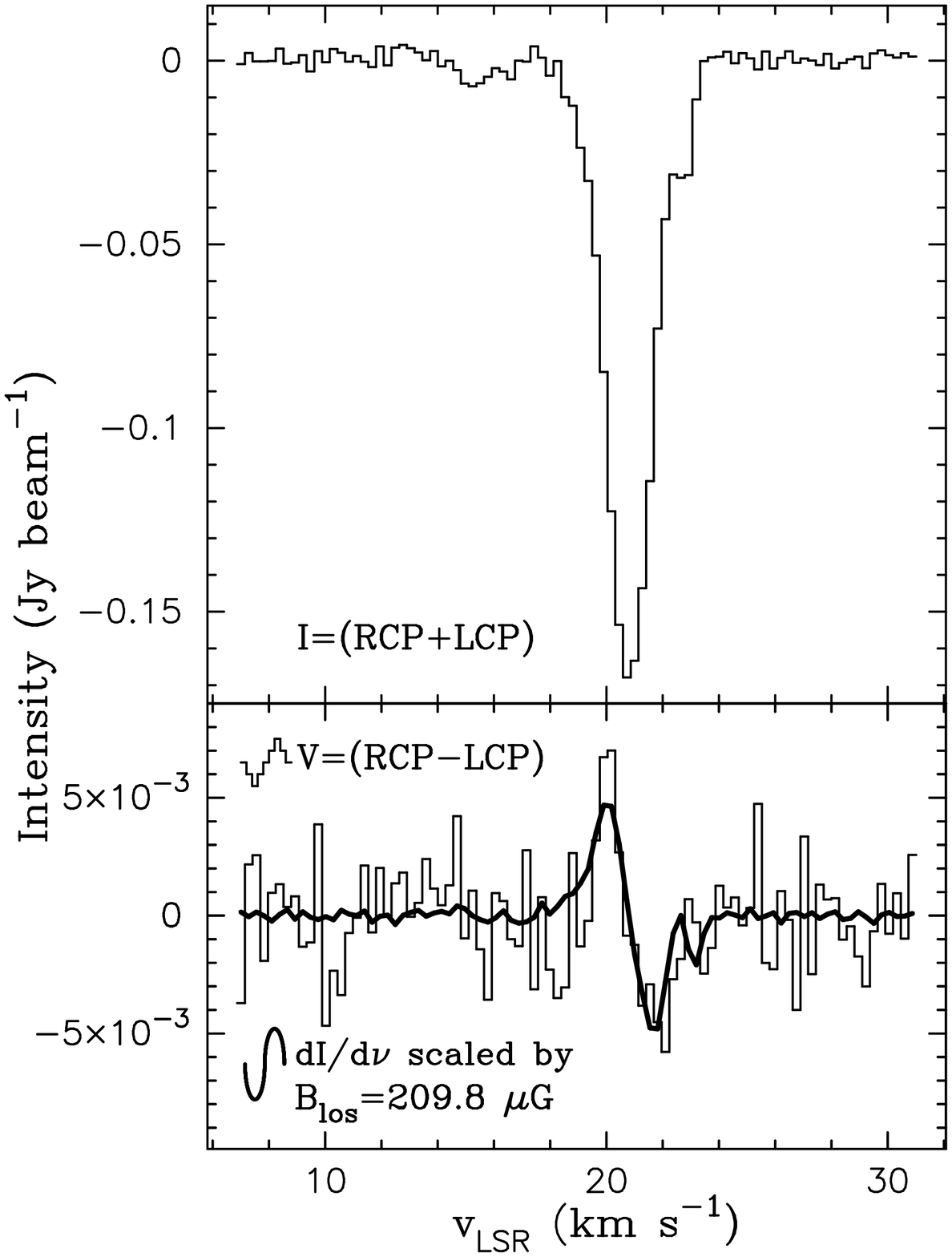}
\end{minipage} \\
\caption{Stokes $I$ ($top; histogram$) and $V$ ($bottom; histogram$) 
profiles at 1665 MHz (left) and 1667 MHz (right) toward the position in S88B-1
marked in Fig.~\ref{fBLOS}\ by a ``+''
($\alpha_{2000} = 19^\text{h}46^\text{m}48\fs3, 
\delta_{2000} = 25^\circ12\arcmin48\farcs3$). The curve superposed
on $V$ in the lower frame on the left shows the derivative of $I$ at 1665 MHz
scaled by \Blos\ = $159.4 \pm 18.0\ \mu$G, whereas the curve superposed 
on $V$ in the lower frame on the right shows the derivative of $I$ at 1667 MHz 
scaled by \Blos\ = $209.8 \pm 29.5\ \mu$G.\label{fB1IVD}}
\end{figure}

\begin{figure}	
\centering
\includegraphics[width=0.5\textwidth]{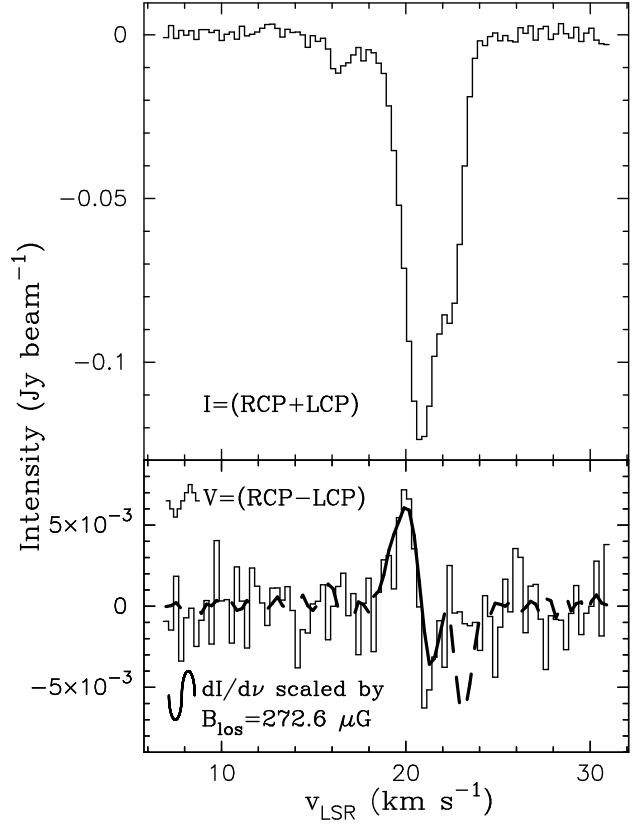}
\caption{Stokes $I$ ($top; histogram$) and $V$ ($bottom; histogram$) 
profiles at 1665 MHz toward the position in S88B-2 
marked in Fig.~\ref{fBLOS}\ by a ``$\times$''
($\alpha_{2000} = 19^\text{h}46^\text{m}49\fs0, 
\delta_{2000} = 25^\circ12\arcmin44\farcs7$). The curve superposed
on $V$ in the lower frame shows the derivative of $I$ at 1665 MHz
scaled by \Blos\ = $272.6 \pm 40.7\ \mu$G; the fit for the field was
made over a restricted range of velocity channels, and the channels over
which the fit was made are indicated by the solid line. \label{fB2IVD}}
\end{figure}

While the velocity structure of OH absorption observed by us toward S88B 
is complicated, the highest values for $N(OH)/T_\text{ex}$ are to the 
north of the continuum peak S88B-2. While this may, in principle, 
be due to variations in $N(OH)$ or $T_\text{ex}$ or both, other tracers 
indicate that the column of absorbing gas is indeed greater in the 
eastern parts of S88B (i.e., toward S88B-2).
Toward the H$\alpha$ peak in the west, 
\citet{fh81}\ found a relatively low \Av\ = 5 mag. Based on their 
observations of IR fine structure lines, \citet{hhb82}\ measured \Av = 
26 $\pm$ 9 mag at the position of the 2.2 NIR $\mu$m peak.
This 2.2 $\mu$m peak was resolved by \citet{dnz2000}\ 
into a system of three stars (82, 83, 84) located on the eastern
edge of the optical nebulosity --- their positions are marked 
in Figure~\ref{fCC}. In contrast, toward S88B-2 in the
east, \citet{ggp03}\ found \Av $\sim$ 70 mag by comparing their 
observed Br$\alpha$ flux with 5 GHz radio continuum data from the literature.
S88B-2 also coincides with the 350 $\mu$m dust continuum peak from \citet{mse02}, as 
shown in Figure~2. There is also an 
extension in the 350 $\mu$m dust continuum image to the northwest, 
which coincides with the location of S88B-1 (Fig.~2). In fact, due to the high
obscuration toward the eastern parts of S88B, a complete census of the 
exciting/ionizing stars of S88B-1 and S88B-2 has proved impossible so far. 
\citet{dnz2000}\ suggested that star 82 could be the ionizing/exciting source of 
S88B-1, based on its high luminosity and central location (see Fig.~\ref{fCC}), but 
admitted the likelihood that other stars also contribute to the excitation/ionization 
in the region. Other possible candidates for exciting sources are the stars L1 (for S88B-2)
and L2 (for S88B-1), detected by \citet{puga04}\ at a near-infrared wavelength of
3.5 $\mu$m (L$^\prime$-band) and marked in Figure~\ref{fCC}; however, the issue remains open because
\citet{puga04}\ could not determine the spectral type of L1 or L2
as they did not detect them in their J, H, or K$^\prime$ near-infrared bands. 

If the absorbing OH gas is thermalized and homogenous (i.e., not 
clumpy within the observing beam), the ratio of the optical depths
of the 1665 MHz and 1667 MHz lines should be 
$\tau_{67}/\tau_{65} = 9/5 = 1.8$. However, the observed profiles 
toward S88B do not show this ratio. Instead, observed ratios 
$\tau_{67}/\tau_{65}$ are in the range 1.0-1.3 over most parts of 
the source. While it is likely that the excitation temperatures
for the two lines at 1665 MHz and 1667 MHz are not the same, it
is unlikely that this difference would cancel out the expected
ratio of optical depths over the whole source. More likely, the
OH gas is clumped on size scales smaller than the beam. Such 
clumpiness has been found also in molecular emission lines.
The CO J=2$-$1 and $^{13}$CO J=2$-$1 and J=1$-$0 data 
(HPBW 22$\arcsec$-33$\arcsec$) of \citet{wf92}\ reveal
a horseshoe-like structure surrounding the optical nebulosity, which
they ascribe to material excited along the periphery of an outflow
cavity that encloses the optical nebula. Indeed, \citet{pm91}\ have 
reported a low collimation bipolar outflow extending over 3$\arcmin$ 
from their CO observations; this CO outflow was also observed by
\citet{rm01}. Both the CO and $^{13}$CO images in \citet{wf92}\
show highly fragmented structure, in agreement with the clumpy
structure suggested by the OH absorption observations. More evidence 
of clumping comes from observations of NH$_3$ lines at high angular 
resolution (HPBW 4$\arcsec$) with the VLA by \citet{goga95}. They 
found compact ammonia structures 
with sizes of 0.2 pc toward both S88B-1 and S88B-2. They suggest 
that these ammonia clumps correspond to compact molecular structures 
embedded within the larger molecular cloud, and that these clumps have been 
heated by the radiation from the star(s) that ionize(s) the associated 
\ion{H}{2}\ regions. Finally, recent C$^{18}$O observations by
\citet{ss07} with a beam HPBW of 15\arcsec\ provide additional 
evidence of clumpy structure (also see \S\ \ref{ssMFE} below); their 
clump B is marked in Figure~\ref{fCC}.  

The OH excitation temperature \Texct\ (which will be required in
the calculations below) cannot be measured based on
absorption studies alone, but reasonable limits for it can be set
from our observations, and a likely value estimated based on 
other tracers. The minimum continuum $S_\nu$ against
which the line is detected is $\sim$ 10 mJy \beam\ --- we can use this
to set an upper limit on \Texct. Using $S_\nu \rightarrow T_{\rm b}:$
1 mJy \beam $\rightarrow$ 17 K, we get an upper limit of 170 K
for \Texct. It is also unlikely that \Texct\ is less than the 
typical dark cloud value of 10 K. Estimates of the kinetic temperature
from other molecular tracers are also relevant if we assume
LTE. From observations of the (2,2) and (3,3) lines of NH$_3$,
\citet{goga95}\ found \subs{T}{kin} = 80 K toward S88B-1
(their western NH$_3$ clump) and 70 K toward S88B-2 (their
eastern NH$_3$ clump). Based on CO observations, \citet{wf92}\ estimated 
\subs{T}{kin}\ $\sim$ 60 K. \citet{ckm86}\ found a dust temperature
for S88B equal to \subs{T}{dust}\ = 42 K. \citet*{mss86}\ fitted a
2-component dust model for S88B, and obtained \subs{T}{dust} = 40 K 
and 116 K. The higher \subs{T}{d}\ may be due to compact embedded sources
or it may reflect an enhanced population of small grains (\citealt{wu89}).
Indeed, \citet{jdm90}\ have reported observations of PAH emission 
features toward S88B. \citet{mse02}\ found T$_\text{dust}$ = 75 K from
their 350 $\mu$m dust continuum observations. We adopt \Texct\ = 60 K for our 
calculations, but note that \Texct\ likely varies
over the source; still, it must lie between 10-170 K.

\subsection{Magnetic Fields}
\label{ssDB}
Our observations of the Zeeman effect in OH absorption lines 
at 1665 and 1667 MHz have yielded \Blos\ values in the range
of 90 to 400 $\mu$G. These values are typical of magnetic
fields in molecular clouds detected via the Zeeman effect in 
the density regimes accessible via \ion{H}{1}\ and OH thermal 
lines, as tabulated in \citet{rmc99}. Note, however, that our
detected fields are higher than the single dish measurements by
\citet{ctk87}, who found \Blos\ = 69 $\pm$ 5 $\mu$G toward 
S88B with the Nancay single dish. This is expected, since the 
low angular resolution (HPBW $3\farcm5 \times 19\arcmin$) 
of Nancay will cause fields to be averaged over a large area, 
leading to much lower \Blos\ values.

\subsubsection{Magnetic Field Energetics}
\label{ssMFE}
The magnetic field in a cloud may consist of a static component $B_{\rm s}$, and
a time-dependent or wave component $B_{\rm w}$. The static component connects the
cloud to the external medium and determines the total magnetic flux throughout
the cloud, but it can only provide support to the cloud perpendicular to the
field lines. The wave component is associated with MHD waves in the cloud,
and it can provide three dimensional support to the cloud. In principle, the observed \Blos\
may consist of contributions from each of these two components. In reality, however,
the Zeeman effect likely traces the strength of the static component of the magnetic
field only. This is because spatial averaging over the beam and along the line 
of sight tends to cancel the contribution from the wave component (e.g.,
\citealt{bt2001}, and references therein). A principal
goal of Zeeman effect measurements, therefore, is to estimate the importance of
the magnetic field to the dynamics and evolution of star forming regions like
S88B. As summarized in \citet{rmc99}, such estimates also require other
physical parameters such as internal velocity dispersion, hydrogen column
density, and radius, to be known. 

The relevant physical parameters mentioned above for estimating the importance 
of the magnetic field are available either from the present observations of 
S88B or from the literature. First, we must decide on an appropriate value 
for the radius. The highest resolution molecular line emission 
observations available to date are the C$^{18}$O observations 
of \citet{ss07}\ referred to above (with beam HPBW 15$\arcsec$). Clumps 
B, C, and D in their observations all have radius $\sim$ 0.2 pc. 
Therefore, we adopt $r = 0.2$ pc for the radius 
in all the following calculations. Next, the hydrogen column density can be 
estimated from the OH line data, subject to certain assumptions. We use an 
average value of $N_{\rm OH}$/\Texct\ = 6.0 $\times$ 10$^{14}$ \cm{-2} K$^{-1}$ 
(see Fig.~\ref{fNOH}). For the excitation temperature, we use \Texct\ = 60 K, 
based on the discussion in \S\ \ref{ssDC}\ above. Then, to get $N_{\rm H_2}$,
we use the conversion ratio $N_{\rm H_2}/N_{\rm OH} = 2 \times 10^6$ from 
\citet{rct95}\ for S106. This makes the OH abundance a factor of 10 greater than 
that in dark clouds found by \citet{c79}. Note, however, that if we consider the
standard conversion ratio $N_{\rm H_2}/A_{\rm v}\ \sim 10^{21}$,
and use $A_{\rm v}$ = 70 mag for S88B (see \S\ \ref{ssDC}\ above), 
we obtain $N_{\rm H_2}/N_{\rm OH} = 1.9 \times 10^6$, which justifies
our use of the \citet{rct95}\ value rather than the dark cloud value; similar
results were found for M17 by \citet{bt2001}.
Using this conversion ratio, we get $N_{\rm H_2} = 7.2 \times 10^{22}$
\cm{-2}. This gives a hydrogen density $n(H_2) = 1.7 \times 10^5$ \cm{-3}, 
which compares well with the value of $n(H_2) = 1.51 \times 10^5$ \cm{-3}
found by \citet{ss07}\ for the clump designated as $B$ in their 
C$^{18}$O observations (see Fig.\ \ref{fCC}). It is worth noting here that at
lower resolution (HPBW 30$\arcsec$), \citet{pm91}\ determined a
mean $n(H_2) \sim 6 \times 10^3$ \cm{-3}. Together, this is further
proof of the clumping in S88B discussed in \ref{ssDC}\ above, 
since the larger beam filling factor in a higher angular resolution observation 
of a clumpy region will naturally give a larger value for the density. Note, 
however, that the density cannot be substantially greater than 
$\sim$10$^5$ \cm{-3}, a conclusion reached by \citet{phi88}\ based on the 
absence of H$_2$CO emission (\citealt{eh81}) toward S88B. Other parameters 
of interest are the velocity dispersion (which can be obtained from the observed 
line width) and the magnetic field; the adopted values for the average observed
line width of OH lines and the average \Blos\ in S88B are listed 
in Table\ \ref{tPARMS}. For the calculations below, we will follow \citet{rmc99}, 
and use total (static) magnetic field strength equal to 2 times \Blos. 

We can now proceed to assess the importance of the magnetic field in S88B,
using a number of approaches. First, we consider the parameter $\beta_p$,
which is the ratio of thermal to magnetic pressures (\citealt{rmc99}); it is a crucial 
parameter in any theory or simulation of the structure and evolution of molecular 
clouds that incorporates magnetic fields. If $\beta_p < 1$, then magnetic fields
are important; the lower the value of $\beta_p$, the more dominant is the 
magnetic field. Based on the adopted values in Table\ \ref{tPARMS}, we find
$\beta_p = 0.14$ in S88B, indicating the importance of the magnetic field
in this region. Our value for $\beta_p$ in S88B is similar to that in high mass star 
forming regions like DR 21 OH 1, where $\beta_p = 0.21$, and W3 (main),
where $\beta_p = 0.13$ (\citealt{rmc99}, and references therein). 
Another way to assess the importance of the magnetic field in S88 is
to use the relation
\begin{equation} \label{eBS}
	B_{S,\ {\rm crit}} = 5 \times 10^{-21}\ N_p\ \quad \mu{\rm G}, 
\end{equation}
where $N_p$ is the average proton column density in the cloud. 
Equation \ref{eBS}\ is a result of equating the static magnetic energy
of the cloud to its gravitational energy; $B_{S,\ {\rm crit}}$ 
is then the average static magnetic field in the cloud that would 
completely support it against self-gravity. If the actual static
magnetic field in the cloud $B_S > B_{S,\ {\rm crit}}$, then the
cloud can be completely supported by the magnetic field, and is 
said to be magnetically subcritical. Further evolution of the 
cloud perpendicular to the field lines will occur primarily via
ambipolar diffusion. The field can be judged to be dynamically important
to the region even if it is less than $B_{S,\ {\rm crit}}$, but is
comparable to it. For $N_{\rm H_2} = 7.2 \times 10^{22}$ \cm{-2}\ from 
Table\ \ref{tPARMS}\ above, we get $B_{S,\ {\rm crit}} = 700 \mu$G.
Our average total (static) magnetic field strength (= 2 times \Blos\ in 
Table\ \ref{tPARMS}) is 400 $\mu$G; that is, the observed magnetic 
field strength is within a factor of $\sim$2 less than the critical field. Therefore, the
magnetic field should be dynamically significant, providing an
important source of support against self gravity. 
Yet another parameter to judge the extent to which a static magnetic 
field can support a cloud against gravitational collapse is the observed
mass to magnetic flux ratio ${[M/{{\Phi}_{\rm B}}]}$, defined by \cite{rmc99}\ in
units of the critical value. This is equivalent to the consideration of $B_{S,\ {\rm crit}}$,
and may be redundant here, but we state it nevertheless to express our results in this
alternative language. If the ratio of observed ${[M/{{\Phi}_{\rm B}}]}$ to the critical
value is greater than 1, then the region is magnetically supercritical, meaning that
the magnetic field cannot by itself provide support against gravitational
collapse. For S88B, we find the observed ${[M/{{\Phi}_{\rm B}}]}$
in units of the critical value to be 1.8. This is consistent with observations to date;
\citet{rmc99}\ found the observed ${[M/{{\Phi}_{\rm B}}]}$ in terms of the critical
value to have an average value equal to 2 in a summary of available Zeeman 
measurements in molecular clouds. The calculated value for S88B indicates 
that whereas the static magnetic field by itself is not enough to support the molecular
cloud against gravitational collapse, it is high enough to provide a significant
source of support. Perhaps more important, however, in a highly supersonic medium like the
S88B molecular cloud, is that for the magnetic field to be important, its energy density must be comparable to, or greater than, 
the turbulent energy density. The supersonic nature of the medium is clear from the value of
the sonic Mach number $m_s$ given in Table 3; from it, we have that 
the ratio of the one-dimensional velocity dispersion ($\sigma$) to the
isothermal sound speed ($c_s$) in S88B is $\sigma/c_s = 2.3$. This is a common result in 
molecular clouds; in his compilation of results for 27 molecular clouds, 
\citet{rmc99}\ found that motions are supersonic by about a factor of 5. Table 3 also lists
the Alfvenic Mach number ($m_A$) to be equal to 1. This implies that the magnetic energy density is 
comparable to the turbulent energy density, that is, these two energies are in approximate equipartition.

Finally, we can use the measured \Blos\ to assess the relative importance of
the gravitational, kinetic, and magnetic energy terms $\cal{W}$, $\cal{T}$, and
$\cal{M}$, in the virial equation (\citealt{mzgh93}). To calculate their values,
we use the expressions for $\cal{W}$, $\cal{T}$, and $\cal{M}$ given in 
\citet{rmc99}; the values are given in Table\ \ref{tVE}. We find that
$\cal{M} \approx {\rm (1/2)}\cal{T} \approx {\rm (1/4)}\cal{W}$, again indicating that 
magnetic fields are a significant, but not the sole, means of support against
gravitational collapse. Finally, we find that
$2 \cal{T} + \cal{M} = {\rm 1.3} \cal{W}$; however, the surface pressure term
in the virial equation, which is not included here, will act in the same sense
as gravity; therefore, S88B is in approximate virial equilibrium.

\section{CONCLUSIONS}
\label{sCONC}
We have observed OH thermal lines in absorption at 1665 and 1667 MHz 
toward the star forming region S88B with the aim of mapping magnetic fields
via the Zeeman effect. The OH absorption line profiles toward this source are
complicated, and contain several blended components toward a number of 
positions. The OH absorbing gas is located in the eastern parts 
of S88B, toward S88B-2 and the eastern section of S88B-1. The ratio of 
optical depths $\tau_{67}/\tau_{65} = 1.0-1.3$ over most parts of the source, 
instead of the value of 1.8 for thermalized and homogenous OH gas. This 
is most likely due to clumping of OH gas, a conclusion supported by 
observations of CO, $^{13}$CO, NH$_3$ and C$^{18}$O emission lines 
from the literature. From the $N(OH)/T_\text{ex}$ plots, we identified 4
prominent OH clumps, which we have designated as OH-a, OH-b, OH-c,
and OH-d. The highest values for $N(OH)/T_\text{ex}$ are to the north of the 
continuum peak S88B-2. Although this could, in principle, be due to an 
enhancement in $N(OH)$ or \Texct\ (or both), it is likely that north of S88B-2
is truly the location of the largest column of absorbing OH gas, as 
supported by independently measured extinction values toward S88B 
from the literature. Based on the observed quantities and adopted values 
of parameters from the literature, we obtain a gas density 
$n(H_2) = 1.7 \times 10^5$ \cm{-3}, which compares well with the 
value of $n(H_2) = 1.5 \times 10^5$ \cm{-3}\ found by \citet{ss07}\ 
for a C$^{18}$O clump coincident with the S88B-1 and S88B-2 sources.

In brief, S88B appears to have an overall geometry similar to M17, in that
there is an edge-on progression from ionized to molecular gas going eastward. 
Our magnetic field map appears to mirror this transition, at least over the 
extended S88B-1 region, where we detected line-of-sight magnetic fields ranging
from 90-400 $\mu$G, with the lowest values to the southwest of the S88B-1 
continuum peak, and the highest values to its northeast. Toward S88B-2, 
the fields are significant at the 3-$\sigma$ level in the 1665 MHz line only,
and are detected over a small region but if we had the 
sensitivity to keep going eastward, the field would likely follow
the same increasing trend. 

The detected fields allow us to assess the importance of the magnetic field
in S88B, using a number of methods. We find that the ratio of thermal to magnetic
pressures, $\beta_p  = 0.14$; a value of $\beta_p < 1$ means that magnetic
fields are important; the lower the value of $\beta_p$, the more dominant the
magnetic field. A second method to assess the importance of the magnetic
field is to compare the observed magnetic field with the critical field that
would be required to completely support the cloud against its 
self-gravity. We find that the observed magnetic field is within a factor of 2
less than the critical field. A third method to assess the importance of the 
magnetic field, equivalent to the second, is to find the ratio of observed mass to magnetic flux in terms 
of the critical value; for S88B, we find this to be 1.8. These considerations
lead us to conclude that the magnetic field in S88B is dynamically significant, 
and provides an important source of support against gravity. Moreover, the magnetic energy
density is comparable to the turbulent energy density, implying that these two
energies are in approximate equipartition, and pointing again to the importance of
the magnetic field in this region.

\acknowledgments
This work has been partially supported by a Cottrell College Science Award
(CCSA) grant from Research Corporation to A.P.S., and by start-up funds 
to A.P.S.\ at DePaul University. T.H.T.\ acknowledges NSF grant AST 03-07642. 
We thank Kaisa Young (\textit{nee} Mueller) and H. Saito for sending us 
digital versions of their data for comparison. We have used extensively 
the NASA Astrophysics Data System (ADS) astronomy abstract service, the 
astro-ph web server, and the SIMBAD database.

\clearpage

\clearpage

%
\begin{deluxetable}{lccccccrrrrr}
\tablenum{1}
\tablewidth{0pt}
\tablecaption{PARAMETERS FOR VLA OBSERVATIONS 
        \protect\label{tOP}}   
\tablehead{
\colhead{Parameter}  &
\colhead{B-configuration}    &
\colhead{C-configuration}}
\startdata
Date    &   2003 Nov 28/29, Dec 1/2    &   1997 Aug 15/16   \nl  
Configuration   &   B     &    C    \nl
R.A. of field center (J2000)   &   $19^{h}46^{m}47^{s}$.7   &
    $19^{h}46^{m}47^{s}$.7    \nl
Decl. of field center (J2000)   &    25$\arcdeg$12$\arcmin$56$\arcsec$.1
    &    25$\arcdeg$13$\arcmin$05$\arcsec$.2   \nl
Total bandwidth (MHz)   &   0.20   &  0.20    \nl
No. of channels   &   128   &   128   \nl
Channel Spacing (km $\mbox{s}^{-1}$)   &   0.28   &   0.28  \nl
Approx. time on source (hr)   &   6.7   &   8.9   \nl
Rest frequency (MHz)   &   1665.402   &   1665.402   \nl
   &  1667.359  &   1667.359   \nl
\enddata
\tablecomments{1$\arcmin$=0.6 pc, for adopted distance to S88B = 2.0 kpc
	(see \S\ \ref{ssISRC})}
\end{deluxetable}

\clearpage

\begin{deluxetable}{lccccrrrrr}
\tablenum{2}
\tablecaption{Source Parameters \protect\label{tPARMS}}   
\tablehead{
\colhead{Parameter}  &&
\colhead{Value} }
\startdata
Radius ($r$)   &&  0.2 pc   \nl
Column Density ($N_{\text{H}_2}$)   &&  $7.2 \times 10^{22}\ \text{cm}^{-2}$  \nl
Particle Density ($n_\text{p}$)   &&   $1.7 \times 10^5\ \text{cm}^{-3}$   \nl
Kinetic Temperature ($T_\text{k}$)   &&  60 K   \nl
Adopted Average FWHM Linewidth ($\Delta v$)   &&   2.5 km s$^{-1}$    \nl
Average Magnetic Field ($B_\text{los}$)   &&  200 $\mu$G  \nl

\enddata

\end{deluxetable}

\clearpage

\begin{deluxetable}{ccccc}
\tablenum{3}
\tablecaption{Source A : Derived Values and Virial Estimates 
	\protect\label{tVE}}   
\tablehead{
\colhead{Parameter}  &
\colhead{Value} }
\startdata
$m_s$   &  4.0   \nl
$m_A $   &  1.0   \nl
$\mbox{$\beta$}_{\rm p}$   &  0.14   \nl
${[M/{{\Phi}_{\rm B}}]}_\text{obs/crit}$  &   1.8   \nl
$\cal{W}$    &  1.3 $\times$  ${\rm 10}^{\rm 46}$  ergs  \nl
$\cal{T}$/$\cal{W}$   &   0.54    \nl
$\cal{M}$/$\cal{W}$   &  0.22  \nl

\enddata
\tablecomments{$m_s$ is the sonic Mach number equal to $\sqrt{3}\, \sigma/c_s$, where $\sigma$ 
is the line velocity dispersion and $c_s$ is the speed of sound; $m_A$ is the Alfvenic Mach number 
equal to $\sqrt{3}\, \sigma$/$v_{\rm A}$, where $v_{\rm A}$ is the Alfv\'{e}n velocity, $\beta_p$ is the ratio 
of thermal to magnetic pressures, $M/{{\Phi}_{\rm B}}$ is the mass to magnetic flux ratio, 
$\cal{W}$ is the virial gravitational energy, $\cal{T}$ is the virial kinetic 
energy, and $\cal{M}$ is the virial magnetic energy; expressions for calculating
all these terms were taken from \citet{rmc99}.}

\end{deluxetable}

\clearpage

\end{document}